\documentclass[sigconf]{acmart}
\AtBeginDocument{%
  }

\copyrightyear{2025}
\acmYear{2025}
\setcopyright{acmlicensed}
\acmConference[CIKM '25]{Proceedings of the 34th ACM International Conference on Information and Knowledge Management}{November 10--14, 2025}{Seoul, Republic of Korea}
\acmBooktitle{Proceedings of the 34th ACM International Conference on Information and Knowledge Management (CIKM '25), November 10--14, 2025, Seoul, Republic of Korea}
\acmDOI{10.1145/3746252.3761137}
\acmISBN{979-8-4007-2040-6/2025/11}
\usepackage{cleveref}
\usepackage{enumitem}
\usepackage{multirow}
\usepackage{hyperref}
\usepackage{makecell}
\usepackage{subfigure}
\newcommand{\name}{Scenario-Wise Rec}
\usepackage[dvipsnames]{xcolor}
\usepackage{bbding}
\usepackage{balance}
\usepackage[misc]{ifsym}

\definecolor{darkgreen}{rgb}{0.0, 0.5, 0.0} 
\begin{document}

%%
%% The "title" command has an optional parameter,
%% allowing the author to define a "short title" to be used in page headers.
\title{\name: A Multi-Scenario Recommendation Benchmark}

%%
%% The "author" command and its associated commands are used to define
%% the authors and their affiliations.
%% Of note is the shared affiliation of the first two authors, and the
%% "authornote" and "authornotemark" commands
%% used to denote shared contribution to the research.

\author{Xiaopeng Li}
\authornote{Both authors contributed equally to the paper.}
\affiliation{%
 \institution{City University of Hong Kong}
 \city{Hong Kong}
 \country{China}
}
\email{xiaopli2-c@my.cityu.edu.hk}

\author{Jingtong Gao}
\authornotemark[1]
\orcid{0000-0002-4470-5972}
\affiliation{%
 \institution{City University of Hong Kong}
 \city{Hong Kong}
 \country{China}
}
\email{jt.g@my.cityu.edu.hk}

\author{Pengyue Jia}
\affiliation{%
 \institution{City University of Hong Kong}
 \city{Hong Kong}
 \country{China}
}
\email{jia.pengyue@my.cityu.edu.hk}

\author{Xiangyu Zhao}
\affiliation{%
 \institution{City University of Hong Kong}
 \city{Hong Kong}
 \country{China}
 }
 \email{xianzhao@cityu.edu.hk}
 \authornote{Corresponding Authors.}

\author{Yichao Wang}
\affiliation{%
 \institution{Huawei Noah’s Ark Lab}
 \city{Shenzhen}
 \country{China}
}
\email{wangyichao5@huawei.com}

\author{Wanyu Wang}
\affiliation{%
 \institution{City University of Hong Kong}
 \city{Hong Kong}
 \country{China}
}
\email{wanyuwang4-c@my.cityu.edu.hk}

\author{Yejing Wang}
\affiliation{%
 \institution{City University of Hong Kong}
 \city{Hong Kong}
 \country{China}
}
\email{yejing.wang@my.cityu.edu.hk}

\author{Yuhao Wang}
\affiliation{%
 \institution{City University of Hong Kong}
 \city{Hong Kong}
 \country{China}
}
\email{yhwang25-c@my.cityu.edu.hk}

\author{Huifeng Guo}
\affiliation{%
 \institution{Huawei Noah’s Ark Lab}
 \city{Shenzhen}
 \country{China}
}
\email{huifeng.guo@huawei.com}

\author{Ruiming Tang$^\dagger$}
\affiliation{%
 \institution{Huawei Noah’s Ark Lab}
 \city{Shenzhen}
 \country{China}
}
\email{tangruiming@huawei.com}

%%
%% By default, the full list of authors will be used in the page
%% headers. Often, this list is too long, and will overlap
%% other information printed in the page headers. This command allows
%% the author to define a more concise list
%% of authors' names for this purpose.
\renewcommand{\shortauthors}{Xiaopeng Li et al.}

%%
%% The abstract is a short summary of the work to be presented in the
%% article.
\begin{abstract}
  Multi-Scenario Recommendation (MSR) tasks, referring to building a unified model to enhance performance across all recommendation scenarios, have recently gained considerable attention. However, current research in MSR faces two significant challenges that hinder the field's development: the absence of uniform procedures for multi-scenario dataset processing, thus hindering fair comparisons, and most models being closed-source, which complicates comparisons with current SOTA models. Consequently, we introduce our benchmark, Scenario-Wise Rec, which comprises six public datasets and twelve baseline models, along with a training and evaluation pipeline. We further validate Scenario-Wise Rec on an industrial advertising dataset, underscoring its robustness. We hope the benchmark will give researchers clear insights into prior work, enabling them to develop novel models and thereby fostering a collaborative research ecosystem in MSR. Our source code is publicly available\footnote{\url{https://github.com/Applied-Machine-Learning-Lab/Scenario-Wise-Rec}}.
\end{abstract}

%%
%% The code below is generated by the tool at http://dl.acm.org/ccs.cfm.
%% Please copy and paste the code instead of the example below.
%%
\begin{CCSXML}
<ccs2012>
   <concept>
       <concept_id>10002951</concept_id>
       <concept_desc>Information systems</concept_desc>
       <concept_significance>500</concept_significance>
       </concept>
   <concept>
       <concept_id>10002951.10003317.10003347.10003350</concept_id>
       <concept_desc>Information systems~Recommender systems</concept_desc>
       <concept_significance>500</concept_significance>
       </concept>
 </ccs2012>
\end{CCSXML}

\ccsdesc[500]{Information systems}
\ccsdesc[500]{Information systems~Recommender systems}

%%
%% Keywords. The author(s) should pick words that accurately describe
%% the work being presented. Separate the keywords with commas.
\keywords{Multi Scenario Recommendation, Recommendation Systems, CTR Prediction}

%% A "teaser" image appears between the author and affiliation
%% information and the body of the document, and typically spans the
%% page.2025-CIKM-xp-jt-Benchmark 

%%
%% This command processes the author and affiliation and title
%% information and builds the first part of the formatted document.
\maketitle

\section{Introduction}

Recommender systems, deeply integrated into the digital world, play a crucial role in mitigating data overload and personalizing user experiences across diverse online platforms~\cite{zhang2019deep, fan2022comprehensive, zhang2021deep, jia2025joint, liu2025bridge, wang2025star, yang2025multi, yi2025adaptive,liu2025contrastive, wang2024gprec, liu2023multi}. Current recommender systems leverage user profiles, behavior sequences, and contextual features to produce customized recommendations for specific user and item scenarios~\cite{zhou2019deep, wang2023single, li2023towards, li2022gromov, liu2025multi, fan2021attacking}. In the face of varied real-world applications, there is a growing body of research on developing models capable of managing multiple recommendation scenarios simultaneously, referred to as the Multi-Scenario Recommendation (MSR) task. MSR models, tailored to unique user and item scenarios, dynamically learn to transfer knowledge across scenarios (also referred to as ``domains'' in some research). This strategy not only addresses data scarcity in less populated scenarios but also enhances overall recommendation performance~\cite{feng2020fusion, xie2022contrastive}.

\begin{figure}[t]
    \centering
    \includegraphics[width=0.9\linewidth]{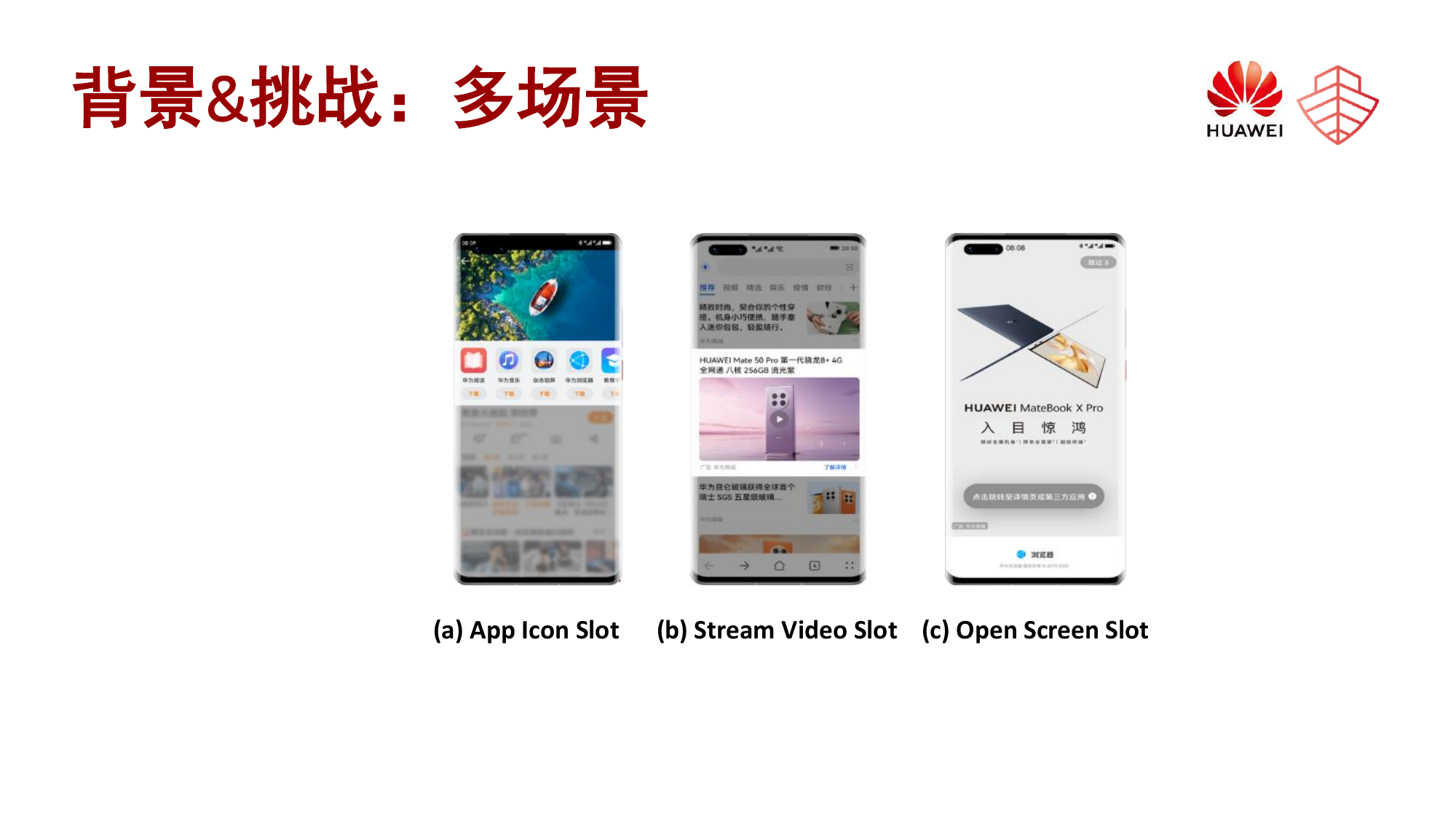}
    \vspace{-1mm}
    \caption{An MSR example in business application: multi-scenario advertising recommendations from real world. Each slot is treated as a specific scenario in modeling.}
    \label{fig:scenario}
    \vspace{-3mm}
\end{figure}

Specifically, multi-scenario recommendation systems aim to develop a unified model capable of generating recommendations across diverse scenarios~\cite{wen2025measure, sheng2021one, yang2022adasparse, wang2022causalint, zhang2024mdmtrec}. These scenarios typically correspond to distinct, predefined domains, such as various advertising sectors, product pages, or manually defined business units, as illustrated in Figure~\ref{fig:scenario}. The primary goal of such models is to leverage knowledge transfer between scenarios to enhance performance within each specific scenario. A key challenge for these models is effectively balancing shared and scenario-specific information, which is crucial for improving overall predictive accuracy. This balance becomes particularly important in real-world applications, where businesses often encounter the challenge of executing recommendation tasks across multiple scenarios~\cite{zhang2022leaving}.

With the development of deep recommender systems~\cite{zhang2019deep, batmaz2019review,wang2023plate} and cross-domain studies~\cite{zhu2021cross, gao2023autotransfer, jia2024d3, wang2024diff, liu2024multifs,fan2023adversarial, gao2023autotransfer}, there has been rapid growth of Multi-Scenario Recommendation methods. Many models, such as STAR~\cite{sheng2021one}, AdaSparse~\cite{yang2022adasparse}, and Causalint~\cite{wang2022causalint}, among others, have been proposed and effectively implemented. However, there is still a lack of a widely recognized benchmark in this area, which poses significant challenges: Firstly, there is a lack of a standardized pipeline for scenario data processing, model training, and model performance evaluation to make fair comparisons between models. Secondly, many current MSR models are closed-source due to corporate privacy protection policies, which complicates reproducibility for researchers, thereby impeding the field's progression in multi-scenario recommendations.

Given these challenges, the need for a well-defined benchmark specifically tailored to multi-scenario recommendations is becoming increasingly urgent. This benchmark should offer standardized procedures for data processing, evaluation, and model interfaces, thereby establishing consistent research norms. In this paper, we introduce \textbf{\name}, the first benchmark exclusively designed for Multi-Scenario Recommendations (MSR). Our benchmark integrates data preprocessing and evaluation protocols for six publicly available datasets, providing a structured framework for model comparison and ensuring fair evaluation conditions. We have developed a standardized model interface and reproduced twelve widely recognized MSR models. To assess the applicability and robustness of our benchmark, we have also applied it to an industrial dataset collected from an advertising platform, demonstrating its real-world performance. Our comprehensive approach not only enables researchers to extract valuable insights from existing MSR work but also fosters a collaborative research environment within this field. The main contributions of this paper are summarized as follows:

\begin{itemize}[leftmargin=*]
\item To the best of our knowledge, \name~is the first benchmark specifically designed for state-of-the-art MSR research, integrating the latest models and a diverse range of MSR datasets. It serves both academic and industrial research communities, facilitating the convergence of advancements from these domains.
\item Our benchmark provides a unified pipeline for MSR tasks, encompassing data preprocessing, model training, and evaluation. It integrates six public datasets and twelve widely recognized MSR models, ensuring fair comparisons and reproducibility. Additionally, the benchmark is validated using an industrial advertising dataset, enhancing its credibility and real-world applicability.
\item We have publicly released our benchmark to facilitate MSR experimentation, allowing researchers to conduct studies more efficiently and derive meaningful insights. This initiative aims to streamline research, foster collaboration, and accelerate progress within the MSR community.
\end{itemize}

\section{Related Work}
Personalization within a single scenario has long been an active research topic~\cite{li2023agent4ranking, jia2024altfs, li2025survey, xu2025align, gao2025samplellm}, focusing primarily on user-item interactions and the underlying patterns between them. However, with the increasing complexity of online platforms, recent years have witnessed a growing interest in multi-scenario recommendation tasks. This trend is fueled by the rapid expansion of user bases and web content. To meet diverse recommendation needs—such as varying advertising slots—platform providers now segment users and content into distinct scenarios, resembling a multi-task learning framework. In response, researchers have begun investigating cross-scenario transfer techniques to effectively address the resulting challenges.
Notable efforts employ Mixture-of-Experts (MoE) structures to manage scenario diversity. Mario~\cite{tian2023multi} captures scenario information through feature scaling modules and dynamically employs MoE structures. HiNet~\cite{zhou2023hinet} uses hierarchical structures for effective scenario information extraction while preserving scenario-specific features. PEPNet~\cite{chang2023pepnet} employs gating units for bottom-level inputs processing and introduces EPNet for scenario feature selection and PPNet for integrating multi-task information. Other approaches address scenario modeling differently. STAR~\cite{sheng2021one} introduces a unified model with scenario-specific and scenario-shared towers to capture unique and shared information. SAR-Net~\cite{shen2021sar} and SAML~\cite{chen2020scenario} use attention mechanisms for scenario feature modeling, facilitating knowledge transfer and improving performance. ADL~\cite{li2023adl} distinguishes scenario communities through an adaptation module, and other work explores scenario knowledge transfer via embedding alignment. CausalInt~\cite{wang2022causalint} uses causal inference for multi-scenario recommendations, and AdaSparse~\cite{yang2022adasparse} applies pruning strategies for scenario adaptation.

Recent studies include HAMUR~\cite{li2023hamur}, which utilizes scenario adapters for improved distribution adaptation, and PLATE~\cite{wang2023plate}, which employs prompt technology for scenario adaptation. D3~\cite{jia2024d3} focuses on autonomous scenario-splitting, while MDRAU~\cite{ju2024multi} leverages ``seen'' scenarios to address ``unseen'' ones. HierRec~\cite{gao2024hierrec} utilizes hierarchical structure for modeling. M-scan~\cite{zhu2024m} introduces a Scenario-Aware Co-Attention mechanism and a Scenario Bias Eliminator. Additionally, Uni-CTR~\cite{fu2023unified} uses LLMs to extract semantic representations across scenarios in MSR, and M$^3$oE~\cite{zhang2024m3oe} refines Mixture-of-Experts (MoE) modules, extending them for multi-scenario and multi-task settings. MLoRA~\cite{yang2024mlora} applies the LoRA module directly for multi-scenario CTR prediction. Our benchmark \name~systematically organises this line of research by offering a unified pipeline that covers datasets, models, training procedures and evaluation protocols, thereby providing researchers with a solid foundation for further exploration of MSR.
\section{Pipeline}
\begin{figure*}[t]
    \centering
    \includegraphics[width=0.75\linewidth]{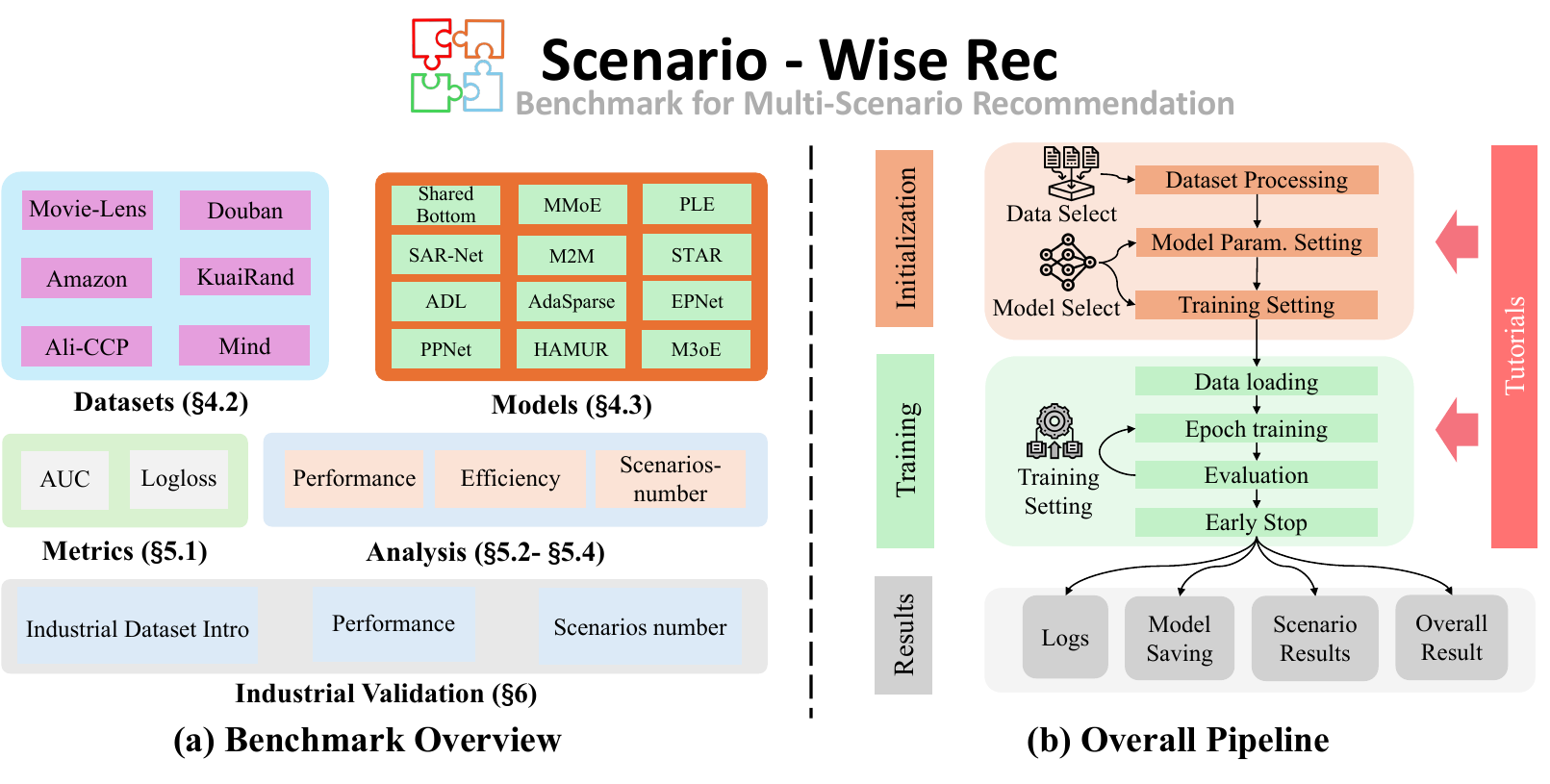}
    \vspace{-3mm}
    \caption{Overall pipeline of \name.}
    \vspace{-2mm}
    \label{fig:overview}
\end{figure*}

In this section, we provide a detailed introduction to the components of our benchmark, whose overall framework is shown in Figure~\ref{fig:overview}.

\begin{itemize}[leftmargin=*]

\item \textbf{Task: Multi-scenario Click-Through Rate Prediction.}  
Our benchmark focuses on Click-Through Rate (CTR) prediction in a multi-scenario setting. In standard CTR prediction~\citep{guo2017deepfm}, the CTR value $\hat{y}$ is predicted by a model $\mathcal{F}_{\theta}$, which takes input features $\boldsymbol{x}$ (e.g., user, item, and context features). This is expressed as $\hat{y} = \mathcal{F}_{\theta}(\boldsymbol{x})$. However, in multi-scenario settings, the input features are extended to include scenario-specific features $\boldsymbol{x}_s$ and a scenario indicator $s \in \{1, ..., S\}$, which identifies the scenario to which the input belongs. Additionally, when designing a multi-scenario model $\mathcal{F}_{\theta^M}$, both scenario-specific and shared features must be jointly considered within the parameter $\theta^M$ across all $S$ scenarios. Mathematically, this is formulated as:
\begin{equation} 
\hat{y} = \mathcal{F}_{\theta^M}(\boldsymbol{x}_g, \boldsymbol{x}_s, s), \quad s \in \{1, ..., S\}
\end{equation}
where $\boldsymbol{x}_g$ denotes the general (scenario-independent) features, $\boldsymbol{x}_s$ represents the scenario-specific features for each scenario $s$, and $\hat{y}$ refers to the CTR prediction.

\item \textbf{Open Datasets.} 
Open datasets play a critical role in research on recommender systems. Even though many datasets are available, their inconsistent usage across MSR studies often impedes fair comparisons. Our proposed benchmark addresses this issue by providing a unified data loading interface, ensuring standardized access to datasets. Specifically, we offer open datasets that have been tested and evaluated within our benchmark. This interface is also designed for easy extensibility, facilitating integration of additional datasets for future experiments (see Section~\ref{sec:Dataset}).

\item \textbf{General Data-Processing Methods.} Inconsistent results across studies often arise from variations in data processing methods. Many studies employ custom approaches but fail to share processed data or detailed procedures, which hampers data reuse. To address this issue, we propose a reproducible data-processing paradigm supporting multiple scenarios, ensuring fair comparisons and repeatable experiments. We implement standardized processing methods, such as scenario feature declaration and common feature filtering, enabling the community to conduct diverse research with consistent data processing practices.

\item \textbf{Unified Model Interface.}\label{sec:modelinterface} Open-source models are often made available through authors' publications or reproductions, but inconsistencies in code and implementation can lead to inconsistent in output. Our benchmark standardizes the modules with a consistent model setup and interface, ensuring reproducible implementations and fair comparisons under unified hyper-parameter settings. We have implemented twelve state-of-the-art models for multi-scenario recommendations, tested on six widely used public datasets and one industrial dataset, demonstrating the effectiveness of this unified interface.

\item \textbf{Training.} We have implemented a unified model training procedure to ensure fair comparisons and scalability. This procedure standardizes the training process, enabling easy extension with various models and datasets. Additionally, we provide functions for saving logs to ensure clear record-keeping of training details and facilitating the reproducibility of experiments.

\item \textbf{Evaluation.} Evaluation metrics are essential for assessing model performance. The use of different metrics across studies complicates fair comparisons. To address this, building on previous works~\citep{sheng2021one, yang2022adasparse, li2023hamur, wang2023plate, chang2023pepnet}, we adopt AUC and Logloss, the two most commonly used metrics, to evaluate model performance across different scenarios. Additionally, we provide a consistent evaluation interface for all models, ensuring fair comparisons.

\item \textbf{Savable Logs \& Settings \& Tutorial.} We offer a unified interface for hyperparameter configuration to standardize evaluation processes and ensure reproducibility. These configurations, along with training logs, are stored in files, enabling users to monitor performance and easily replicate results. To further assist researchers, especially newcomers, we provide a comprehensive tutorial covering environment setup, dataset acquisition, preprocessing, model training, and evaluation. Additionally, we introduce custom-designed MSR models and datasets, supporting personalized model development.

\end{itemize}

\section{Benchmarking for Multi-Scenario Recommendation}\label{sec:dataset}
This section presents a comprehensive overview of our benchmark, including comparisons with existing benchmarks, as well as the datasets and multi-scenario baselines used in our study.

\subsection{Comparison with Existing Benchmarks}

\begin{table}[t]
\renewcommand{\arraystretch}{0.8} %可以控制行间距
  \centering
  \vspace{-2mm}
  \caption{Comparison with existing benchmarks.}
  \vspace{-2mm}
\setlength{\tabcolsep}{1mm}{
\scalebox{0.7}{
    \begin{tabular}{lccccc}
    \toprule
    \textbf{Benchmark}  & \textbf{Industrial Validation} & \textbf{Tutorial}  & \textbf{Custom Settings} & \textbf{Task} & \textbf{Year} \\
    \midrule
    Spotlight~\cite{kula2017spotlight}      & {\color{red}\XSolidBrush}     & {\color{darkgreen}\Checkmark}     &  {\color{red}\XSolidBrush}    & Multiple & 2017 \\
    DeepCTR~\cite{shen2017deepctr}    & {\color{red}\XSolidBrush}     & {\color{darkgreen}\Checkmark}     & {\color{darkgreen}\Checkmark}   & CTR     & 2017 \\
    RecBole~\cite{recbole[1.0]}    & {\color{red}\XSolidBrush}     & {\color{darkgreen}\Checkmark}    & {\color{darkgreen}\Checkmark}       & Multiple & 2021 \\
    FuxiCTR~\cite{zhu2021open}    & {\color{red}\XSolidBrush}     & {\color{darkgreen}\Checkmark}    & {\color{darkgreen}\Checkmark}    & CTR    & 2021 \\
    RecBole-CDR~\cite{recbole[2.0]}    & {\color{red}\XSolidBrush}     & {\color{red}\XSolidBrush}    & {\color{red}\XSolidBrush}       & CDR & 2022\\
    SELFRec~\cite{yu2023self}    & {\color{red}\XSolidBrush}     & {\color{red}\XSolidBrush}     & {\color{darkgreen}\Checkmark}       & SRS & 2023 \\
    MMLRec~\cite{yuan2024mmlrec} & {\color{red}\XSolidBrush} & {\color{red}\XSolidBrush} & {\color{red}\XSolidBrush}  & MTMS-CTR & 2024\\
    \midrule
    \textbf{Scenario-Wise Rec} & {\color{darkgreen}\Checkmark}& {\color{darkgreen}\Checkmark} & {\color{darkgreen}\Checkmark}  & MS-CTR & \textbf{2025}\\
    \bottomrule 
    \vspace{-5mm}
    \end{tabular}%
    
  \label{tab:comparebaseline}%
  }}
  % \vspace{-5mm}
\end{table}%

We summarize the relevant benchmarks in Table~\ref{tab:comparebaseline}. Compared to Spotlight~\cite{kula2017spotlight}, DeepCTR~\cite{shen2017deepctr}, RecBole~\cite{recbole[1.0]}, FuxiCTR~\cite{zhu2021open}, RecBole-CDR~\cite{recbole[2.0]}, and SELFRec~\cite{yu2023self}, our benchmark is explicitly designed for multi-scenario CTR tasks. It encompasses twelve MSR models and six datasets, thereby significantly extending the scope and specialization of existing benchmarks. Moreover, unlike prior work, our benchmark offers industrial validation, comprehensive tutorials, and customizable settings, including the construction of custom multi-scenario datasets and models. Because it is exclusively focused on MSR, the benchmark introduces a specialized data processing pipeline and integrates a broader set of MSR-specific datasets and models, providing more comprehensive and practical insights into MSR learning.

\subsection{Datasets} \label{sec:Dataset}
In line with the principles of fair comparison and ease of use, our benchmark selects widely used, multi-scenario open datasets that vary in feature numbers and data volumes. Specifically, for public datasets, we include MovieLens, KuaiRand, Ali-CCP, Amazon, Douban, and Mind. The dataset statistics are listed in Table~\ref{tab:dataset}. The discussion on scenario splitting strategies and scenario intersection analysis can be found in Appendix~\ref{sec_appendix:Scenario Information Analysis}.

\begin{table*}[h]
\renewcommand{\arraystretch}{0.75} %可以控制行间距
    \centering
    \vspace{-2mm}
    \caption{Dataset statistics for each scenario. $^\dagger$ indicates only part of scenarios are shown. }
    \vspace{-2mm}
    \label{tab:dataset}
    \setlength{\tabcolsep}{0.8mm}{
    \scalebox{0.9}{
    \begin{tabular}{c|ccc|ccccccccc}
\toprule
         & \multicolumn{3}{c|}{\textbf{MovieLens}} & \multicolumn{5}{c|}{\textbf{KuaiRand}} & \multicolumn{4}{c}{\textbf{Mind}} \\ 
        \midrule
        Scenario Index &  S-0 &  S-1 &  S-2 &  S-0 &  S-1 &  S-2 &  S-3 & \multicolumn{1}{c|}{ S-4} &  S-0 &  S-1 &  S-2 &  S-3 \\ 
        \midrule
        \# Interaction & 210,747 & 395,556 & 393,906 & 2,407,352 & 7,760,237 & 895,385 & 402,366 & \multicolumn{1}{c|}{183,403} & 26,057,579 & 11,206,494 & 10,237,589 & 9,226,382 \\ 
        \# User & 1,325 & 2,096 & 2,619 & 961 & 991 & 171 & 832 & \multicolumn{1}{c|}{832} & 737,687 & 678,268 & 696,918 & 656,970 \\ 
        \# Item & 3,429 & 3,508 & 3,595 & 1,596,491 & 2,741,383 & 332,210 & 547,908 & \multicolumn{1}{c|}{43,106} & 8,086 & 1,797 & 8,284 & 1,804 \\ 
        
        \bottomrule
        \toprule
         & \multicolumn{3}{c|}{\textbf{Douban}} & \multicolumn{3}{c|}{\textbf{Ali-CCP}} & \multicolumn{3}{c|}{\textbf{Amazon}} & \multicolumn{3}{c}{\textbf{Industrial}$^\dagger$} \\ 
         \midrule
        Scenario Index &  S-0 &  S-1 & \multicolumn{1}{c|}{ S-2} &  S-0 &  S-1 &  \multicolumn{1}{c|}{ S-2} &  S-0 &  S-1 & \multicolumn{1}{c|}{ S-2} &  S-0 &  S-1 &  S-2 \\ 
        \midrule
        \# Interaction & 227,251 & 179,847 & 1,278,401 & 32,236,951 & 639,897 & \multicolumn{1}{c|}{52,439,671} & 198,502 & 278,677 & \multicolumn{1}{c|}{346,355} & 301,654 & 91,468 & 22,986 \\ 
        
        \# User & 2,212 & 1,820 & 2,712 & 89,283 & 2,561 & \multicolumn{1}{c|}{150,471} & 22,363 & 39,387 & \multicolumn{1}{c|}{38,609} & - & - & - \\ 
        
        \# Item & 95,872 & 79,878 & 34,893 & 465,870 & 188,610 & \multicolumn{1}{c|}{467,122} & 12,101 & 23,033 & \multicolumn{1}{c|}{18,534} & - & - & - \\ 
        \bottomrule
    \end{tabular}
    }}
    \vspace{-2mm}
\end{table*}

\begin{itemize}[leftmargin=*]
\item 
\textbf{MovieLens}~\cite{harper2015movielens}: 
The MovieLens dataset is a widely used collection of movie ratings and associated information for recommender tasks. It includes user ratings, demographic data, movie metadata, and user preferences, comprising 1 million anonymous ratings for approximately 4,000 movies made by 6,000 users. In this benchmark, we follow prior works~\cite{zhang2024m3oe, li2023hamur} in partitioning the dataset into three distinct groups based on the "age" feature: ``1-24'', ``25-34'', and ``35+''. 

\item 
\textbf{KuaiRand}~\cite{gao2022kuairand}: The KuaiRand dataset is an unbiased recommendation dataset collected from Kuaishou. It includes 11 million interactions from 1,000 users and 4 million videos. Following previous works like \cite{gao2024hierrec}, different scenarios are defined by the ``tab'' identifier, which represents various advertising positions within the app. The values of the ``tab'' identifier range from 0 to 14, indicating the locations where interactions occurred. For training and testing, we extracted data from the top five scenarios with the most interactions.

\item
\textbf{Ali-CCP}~\cite{ma2018entire}: Ali-CCP is a large-scale CTR recommendation dataset gathered from real-world traffic logs of the recommender system on Taobao, one of the largest online retail platforms globally. In this dataset, we follow previous works~\cite{wang2022causalint, li2023adl}, splitting the scenarios based on the feature named ``301'', which indicates the position where the click occurred.

\item
\textbf{Amazon}~\cite{cui2020herograph}: The Amazon 5-core dataset, a widely used resource for CTR prediction, contains records of user interactions on the Amazon shopping platform. In our benchmark, we treat different categories as distinct scenarios like previous works~\cite{fu2023unified, yang2024mlora}. Specifically, three scenarios, ``Clothing'', ``Beauty'', and ``Health'' are selected for training and evaluation.
\item
\textbf{Douban}~\cite{zhu2020graphical}: The Douban dataset, a real-world collection derived from the Douban platform, is divided into three subsets: Douban-book, Douban-music, and Douban-movie. All subsets share the same users, with each platform treated as a distinct scenario. User features like ``living place'' and ``user ID'' are retained. Following previous works \cite{zhu2020graphical, wang2023plate}, ratings above 3 are considered positive labels, and those below 3 are treated as negative.

\item
\textbf{Mind}~\cite{wu2020mind}: The MIND dataset, designed for news recommendation , is gathered from the Microsoft News platform. In our benchmark, we collect metadata from both the training and validation datasets of MIND for experimentation. We retain item features ``category'' and ``subcategory'', with user clicks considered positive and non-clicks as negative. Scenarios are split based on genres, specifically, the four largest genres, ``news'', ``lifestyle'', ``sports'', and ``finance'' are treated as distinct scenarios. This configuration includes 748 million users, more than 20,000 items, and over 56 million interactions.

\end{itemize}

\subsection{Models}\label{sec:Multi-Scenario Recommendations model}

Our benchmark selects 12 widely recognized MSR models for an extensive comparison. A detailed introduction to these models is provided as follows:

\begin{itemize}[leftmargin=*]

\item
\textbf{Shared Bottom}~\cite{caruana1997multitask}:
The Shared Bottom model is an approach for multi-task recommendation tasks. It learns a shared representation with a shared network to capture the latent patterns. Afterward, different network towers are applied to different tasks for task-specific modeling. In MSR, it has also been applied as a commonly used baseline by treating different scenarios as different recommendation tasks~\cite{sheng2021one,wang2022causalint}.

\item
\textbf{MMoE}~\cite{ma2018modeling}:
The Multi-gate Mixture-of-Experts (MMoE) model is a widely adopted approach for multi-task learning. It utilizes multiple expert networks as foundational structure, along with several gating networks that regulate the connections between the experts. By explicitly modeling the relationships between tasks, MMoE delivers enhanced performance. Similar to other multi-task models, MMoE can be extended to multi-scenario recommendations by treating distinct scenarios as separate recommendation tasks.

\item
\textbf{PLE}~\cite{tang2020progressive}:
The Progressive Layered Extraction (PLE) model is another effective method for multi-task learning in recommender systems. PLE explicitly separates shared components from task-specific components and employs a progressive routing mechanism to progressively extract deeper semantic knowledge. This approach has significantly outperformed state-of-the-art multi-task learning models across various domains. Similarly, PLE can also be applied as a multi-scenario recommendation (MSR) model by treating distinct scenarios as separate recommendation tasks.

\item
\textbf{STAR}~\cite{sheng2021one}:
The Star Topology Adaptive Recommender (STAR) model tackles the challenge of CTR prediction for MSR within large-scale commercial platforms. It facilitates multi-scenario learning by sharing a central network that captures the shared patterns across scenarios, alongside scenario-specific networks tailored to each individual scenario. During the inference stage, the weights of the shared network and the scenario-specific networks are multiplied for each scenario. Online validation has demonstrated the effectiveness of STAR, with notable improvements in both CTR and Revenue Per Mille (RPM) observed after deployment in Alibaba's display advertising system.

\item
\textbf{SAR-Net}~\cite{shen2021sar}:
The Scenario-Aware Ranking Network (SAR-Net), employs two attention modules to learn cross-scenario user interests and a scenario-specific transformation layer to extract relevant features. Additionally, SAR-Net incorporates debiasing expert networks to mitigate bias and a Fairness Coefficient to correct for manual interventions. offline results and online A/B testing validates the effectiveness of SAR-Net, which has been successfully deployed to support hundreds of travel scenarios on Alibaba's online travel marketing platform.

\item
\textbf{M2M}~\cite{zhang2022leaving}:
The Multi-Scenario Multi-Task Meta-Learning~(M2M) model is a novel approach designed to address the challenges of multi-task multi-scenario advertiser modeling in e-commerce platforms. It leverages a backbone network to learn advertiser and task representations and incorporates a Meta Unit to learn scenario-specific knowledge. A Meta Learning Mechanism, with meta attention and meta residual layers, helps capture inter-scenario correlations and improves scenario-specific feature representations. During our benchmark, we set the number of the meta-towers to 1 to correspond to the single CTR prediction task.

\item
\textbf{AdaSparse}~\cite{yang2022adasparse}:
AdaSparse is designed for multi-scenario CTR prediction and aims to adaptively learn the sparse structures of scenario models. Specifically, AdaSparse introduces a lightweight net as a pruner, operating scenario-pruning process for each layer within individual scenario towers. During pruning, novel fusion strategies are employed, combining binary and scale approaches to enhance pruning performance, effectively eliminating as much redundant information as possible. The results show significant improvements in both public datasets and online A/B tests within Alibaba's advertising system's CTR platform.

\item
\textbf{ADL}~\cite{li2023adl}:
The Adaptive Distribution Learning Framework (ADL), a novel multi-distribution method, concentrates on multi-scenario CTR prediction. It features an end-to-end, hierarchical structure that includes a clustering process and a classification process. The core component, the distribution adaptation module, employs a routing mechanism, adaptively determining the distribution cluster for each sample. This model effectively captures the commonalities and distinctions among various distributions, thereby enhancing the model's representation capability without relying on prior knowledge for predefined data allocation. Extensive experiments are conducted on public datasets, and an on an industrial dataset from Alibaba's online system consisting of 10 distinct scenarios. The results demonstrate its effectiveness and efficiency compared to other models.
\begin{table*}[!h]
\centering
\renewcommand{\arraystretch}{0.9} 
\setlength{\tabcolsep}{2mm}{
  \caption{Performance comparison. The best results are in \textbf{bold}. The next best results are  \underline{underlined}. $\pm$ indicates standard error. $\uparrow$ means higher is better, $\downarrow$ means lower the better. ``*'' indicates statistical significance (i.e. two-sided t-test with $p<0.05$).}
  \vspace{-2mm}
  \label{tab:Performance}
  \scalebox{0.9}{
    \begin{tabular}{c|cc|cc|cc}
    \toprule
    \multirow{2}[2]{*}{\textbf{Model}} & \multicolumn{2}{c|}{\textbf{MovieLens}} & \multicolumn{2}{c|}{\textbf{KuaiRand}} & \multicolumn{2}{c}{\textbf{Ali-CCP}} \\
    & AUC$\uparrow$ & Logloss$\downarrow$ & AUC$\uparrow$ & Logloss$\downarrow$ & AUC$\uparrow$ & Logloss$\downarrow$  \\
    \midrule
    SharedBottom & 0.8095 $_{\pm0.0018}$  & 0.5228 $_{\pm0.0016}$  & 0.7793 $_{\pm 0.0009}$  & 0.5483 $_{\pm 0.0010}$  & 0.6232 $_{\pm 0.0021}$  & 0.1628 $_{\pm 0.0012}$   \\
    MMoE  & 0.8086 $_{\pm0.0020}$  & 0.5218 $_{\pm0.0016}$  & 0.7794 $_{\pm 0.0011}$  & 0.5477 $_{\pm 0.0012}$  & 0.6242 $_{\pm 0.0016}$  & 0.1621 $_{\pm 0.0011}$   \\
    PLE  & 0.8091 $_{\pm0.0013}$  & 0.5257 $_{\pm0.0014}$  & 0.7796 $_{\pm 0.0010}$  & 0.5495 $_{\pm 0.0010}$  & 0.6250 $_{\pm 0.0014}$  & 0.1617 $_{\pm 0.0013}$   \\
    STAR   & 0.8096 $_{\pm0.0015}$  & 0.5258 $_{\pm0.0010}$  & 0.7806 $_{\pm 0.0008}$  & 0.5404 $_{\pm 0.0010}$  & \underline{0.6253} $_{\pm 0.0015}$  & 0.1613 $_{\pm 0.0010}$ \\
    SAR-Net & 0.8092 $_{\pm0.0014}$  & 0.5245 $_{\pm0.0010}$  & 0.7816 $_{\pm 0.0010}$  & \textbf{0.5393}$^*$ $_{\pm 0.0010}$  & 0.6245 $_{\pm 0.0016}$  & 0.1616 $_{\pm 0.0010}$   \\
    M2M   & 0.8115 $_{\pm0.0011}$  & 0.5213 $_{\pm0.0013}$  & \textbf{0.7821}$^*$ $_{\pm 0.0012}$  & \underline{0.5397} $_{\pm 0.0010}$  & \textbf{0.6257}$^*$ $_{\pm 0.0014}$  & \textbf{0.1611}$^*$ $_{\pm 0.0011}$ \\
    AdaSparse   & 0.8108 $_{\pm0.0010}$  & \underline{0.5205} $_{\pm0.0010}$  & 0.7816 $_{\pm 0.0011}$  & 0.5399 $_{\pm 0.0010}$  & 0.6239 $_{\pm 0.0020}$  & 0.1614 $_{\pm 0.0012}$ \\
    ADL         & 0.8083 $_{\pm0.0010}$  & 0.5238 $_{\pm0.0010}$  & 0.7773 $_{\pm 0.0008}$  & 0.5436 $_{\pm 0.0009}$  & 0.6233 $_{\pm 0.0015}$  & 0.1619 $_{\pm 0.0012}$ \\
    EPNet       & 0.8097 $_{\pm0.0019}$  & 0.5215 $_{\pm0.0010}$  & 0.7801 $_{\pm 0.0015}$  & 0.5411 $_{\pm 0.0013}$  & 0.6236 $_{\pm 0.0014}$  & \underline{0.1612}$_{\pm 0.0010}$ \\
    PPNet       & 0.8063 $_{\pm0.0012}$  & 0.5257 $_{\pm0.0012}$  & 0.7800 $_{\pm 0.0016}$  & 0.5408 $_{\pm 0.0017}$  & 0.6144 $_{\pm 0.0009}$  & 0.1622 $_{\pm 0.0011}$ \\
    HAMUR       & \textbf{0.8133}$^*$ $_{\pm0.0009}$  & \textbf{0.5193}$^*$ $_{\pm0.0011}$ & \underline{0.7820} $_{\pm0.0015}$& \underline{0.5397} $_{\pm0.0013}$& 0.6235 $_{\pm0.0011} $& 0.1614 $_{\pm0.0010}$  \\
    M$^3$oE & \underline{0.8116} $_{\pm0.0010}$ & 0.5211 $_{\pm0.0008}$ & 0.7812 $_{\pm0.0011}$ & 0.5399 $_{\pm0.0012}$ &  0.6249$_{\pm0.0009}$  & 0.1610 $_{\pm0.0010}$  \\ 
    
    \bottomrule
    \toprule
    
    \multirow{2}[2]{*}{\textbf{Model}} & \multicolumn{2}{c|}{\textbf{Amazon}} & \multicolumn{2}{c|}{\textbf{Douban}} & \multicolumn{2}{c}{\textbf{Mind}}  \\
    & AUC$\uparrow$ & Logloss$\downarrow$ & AUC$\uparrow$ & Logloss$\downarrow$ & AUC$\uparrow$ & Logloss$\downarrow$  \\
    \midrule

    SharedBottom & 0.6792 $_{\pm0.0027}$  & 0.4790 $_{\pm0.0026}$  & 0.7993 $_{\pm 0.0011}$  & 0.5178 $_{\pm 0.0013}$  & \underline{0.7509}$_{\pm 0.0011}$  & \underline{0.1600}$_{\pm 0.0014}$ \\
    MMoE        & 0.6744 $_{\pm0.0025}$  & 0.4963 $_{\pm0.0025}$  & 0.7978 $_{\pm 0.0014}$  & 0.5192 $_{\pm 0.0010}$  & 0.7508 $_{\pm 0.0012}$  & \underline{0.1600}$_{\pm 0.0012}$ \\
    PLE         & 0.6721 $_{\pm0.0020}$  & 0.4945 $_{\pm0.0020}$  & 0.7977 $_{\pm 0.0015}$  & 0.5196 $_{\pm 0.0017}$  & 0.7503 $_{\pm 0.0020}$  & 0.1601 $_{\pm 0.0017}$ \\
    STAR        & 0.6738 $_{\pm0.0022}$  & 0.4966 $_{\pm0.0018}$  & 0.7957 $_{\pm 0.0015}$  & 0.5218 $_{\pm 0.0017}$  & \textbf{0.7512}$^*$ $_{\pm 0.0018}$  & \textbf{0.1593}$^*$ $_{\pm 0.0015}$ \\
    SAR-Net     & 0.7071 $_{\pm0.0026}$  & \textbf{0.4595}$^*$ $_{\pm0.0022}$  & \underline{0.8033} $_{\pm 0.0014}$  & \textbf{0.5131}$^*$ $_{\pm 0.0018}$  & 0.7490 $_{\pm 0.0013}$  & 0.1604 $_{\pm 0.0015}$ \\
    M2M         & 0.6865 $_{\pm0.0023}$  & 0.4943 $_{\pm0.0021}$  & 0.7962 $_{\pm 0.0014}$  & 0.5229 $_{\pm 0.0019}$  & 0.7508 $_{\pm 0.0013}$  & 0.1601 $_{\pm 0.0017}$ \\
    AdaSparse   & 0.6888 $_{\pm0.0020}$  & 0.4831 $_{\pm0.0020}$  & 0.7963 $_{\pm 0.0013}$  & 0.5216 $_{\pm 0.0011}$  & 0.7497 $_{\pm 0.0010}$  & 0.1604 $_{\pm 0.0019}$ \\
    ADL         & \underline{0.7085} $_{\pm0.0030}$  & \underline{0.4658} $_{\pm0.0022}$  & 0.8003 $_{\pm 0.0012}$  & 0.5187 $_{\pm 0.0013}$  & 0.7328 $_{\pm 0.0015}$  & 0.1629 $_{\pm 0.0021}$ \\
    EPNet       & \textbf{0.7101}$^*$ $_{\pm0.0025}$  & 0.4688 $_{\pm0.0024}$  & 0.7997 $_{\pm 0.0014}$  & 0.5182 $_{\pm 0.0010}$  & 0.7418 $_{\pm 0.0017}$  & 0.1616 $_{\pm 0.0018}$ \\
    PPNet       & 0.6791 $_{\pm0.0025}$  & 0.4730 $_{\pm0.0022}$  & 0.7994 $_{\pm 0.0010}$  & 0.5175 $_{\pm 0.0009}$  & 0.7494 $_{\pm 0.0018}$  & 0.1603 $_{\pm 0.0014}$ \\
    HAMUR       & 0.6730 $_{\pm0.0022}$  & 0.4890 $_{\pm0.0019}$ & 0.7979 $_{\pm0.0012}$ & 0.5197 $_{\pm0.0011}$ & 0.7494 $_{\pm0.0015}$ & 0.1603 $_{\pm0.0015}$  \\
    M$^3$oE     & 0.7010 $_{\pm0.0019}$ & 0.4698 $_{\pm0.0018}$ & \textbf{0.8036}$^*$ $_{\pm0.0010}$& \underline{0.5140} $_{\pm0.0009}$&  0.7451 $_{\pm0.0012}$ & 0.1612 $_{\pm0.0011}$  \\ 

    \bottomrule
    \end{tabular}}%
    \vspace{-2mm}
  }
\end{table*}%
\item
\textbf{EPNet \& PPNet}~\cite{chang2023pepnet}:
PPNet and EPNet are two submodels within the Parameter- and Embedding-Personalized Network (PEPNet). EPNet performs personalized embedding selection to fuse features with varying importance for users across scenarios. PPNet modifies the parameters of the deep neural network in a personalized manner to balance targets with varying sparsity for different users across multiple tasks. By leveraging both PPNet and EPNet, PEPNet can effectively handle multi-task recommendations in multi-scenario settings. In \name, we apply these two models in multi-scenario settings, specifically, the number of meta-towers in PPNet is set equal to the number of scenarios to align with the CTR prediction task for each scenario.

\item
\textbf{HAMUR}~\cite{li2023hamur}:
HAMUR employs two kinds of adapters for MSR: domain-specific adapters and a domain-shared hyper-network. The domain-specific adapter is a modular component that can be seamlessly integrated into various recommendation models, enabling flexible adaptations for each domain. The shared hyper-network dynamically generates parameters for these adapters by implicitly capturing shared patterns across domains. Extensive offline experiments demonstrate HAMUR's ability to outperform state-of-the-art models by enhancing predictive accuracy across diverse domains.

\item
\textbf{M$^3$oE}~\cite{zhang2024m3oe}:
The M$^3$oE framework is designed to address challenges across diverse domains and tasks. At its core, M$^3$oE employs three distinct MoE modules, each dedicated to managing domain-specific preferences and task-specific behaviors. Additionally, it integrates a two-level fusion mechanism to effectively combine features across both domains and tasks. The framework's adaptability is further enhanced through the use of AutoML, which dynamically optimizes its structure, enabling efficient cross-domain and cross-task knowledge transfer and ultimately demonstrating superior performance.

\end{itemize}

\begin{table*}[h]
\renewcommand{\arraystretch}{0.8} %可以控制行间距
\setlength{\tabcolsep}{3mm}{
  \centering
  \caption{Efficiency analysis. ``Train'' denotes the average training time per epoch, whereas ``Infer'' denotes inference time per batch on the test set, the batch size is 9,048 for KuaiRand, 102,400 for Ali-CCP and 4,096 for the rest.}
  \vspace{-2mm}
  \label{tab:efficiency}
  \scalebox{0.95}{
    \begin{tabular}{c|ccc|ccc|ccc}
\toprule
    \multirow{2}[2]{*}{\textbf{Model}} & \multicolumn{3}{c|}{{\textbf{MovieLens}}} & \multicolumn{3}{c|}{{\textbf{Ali-CCP}}} & \multicolumn{3}{c}{\textbf{Amazon}} \\
% \hline
% \cmidrule{2-10}          
& Train (s) & {Infer (ms)} & {Params.} & Train (s) & {Infer (ms)} & {Params.} & Train (s) & {Infer (ms)} & {Params.} \\
    \midrule
    {SharedBottom} & 8.68 & {5.49}& {227.59K} & 2918.22 & 29.20 & {25.69M} & 3.09 & {3.61}& {2.22M} \\
    {MMoE} & 9.89 & {5.16}& {217.80K} & 3100.01 & 26.50 & {25.40M} & 4.49 & {4.15}& {2.21M} \\
    {PLE} & 8.17 & {6.16}& {224.20K} & 2559.67 & 29.37 & {25.96M} & 5.57 & {4.25}& {2.22M} \\
    {STAR} & 8.72 & {4.88}& {308.63K} & 2992.08 & 30.99 & {25.54M} & 5.87 & {4.60}& {2.27M} \\
    {SAR-Net} & 7.05 & {7.64}& {239.34K} & 2880.83 & 29.77 & {25.07M} & 4.06 & {3.95}& {2.23M} \\
    {M2M} & 11.71 & {11.83}& {372.53K} & 3042.11 & 28.09 & {26.68M} & 13.59 & {11.71}& {2.31M} \\
    {AdaSparse} & 8.11 & {4.02}& {230.32K} & 2885.73 & 27.70 & {25.33M} & 3.70 & {3.80}& {2.22M} \\    
    {ADL} & 8.54 & {4.18}& {257.49K} & 3194.35 & 28.69 & {25.52M} & 5.86 & {4.49}& {2.24M} \\
    {EPNet} & 8.65 & {4.29}& {232.33K} & 3014.37 & 29.45 & {25.23M} & 4.76 & {3.98}& {2.22M} \\
    {PPNet} & 9.83 & {4.32}& {349.68K} & 2910.49 & 27.11 & {26.23M} & 4.38 & {4.12}& {2.36M} \\
    {HAMUR} & 9.88& 6.96& {362.43K}& 3015.65& 29.23& 27.62M& 5.21& 4.28& 2.38M\\
    {M$^3$oE} & 8.92& 5.85& {296.57K}& 2996.32& 30.02& 25.65M& 4.95& 4.05& 2.27M\\
    
    \bottomrule
    \toprule
    
    \multirow{2}[2]{*}{\textbf{Model}} & \multicolumn{3}{c|}{\textbf{Douban}} & \multicolumn{3}{c|}{\textbf{KuaiRand}} & \multicolumn{3}{c}{\textbf{Mind}} \\
% \cmidrule{2-10}  
% \hline
& Train (s) & {Infer (ms)} & {Params.} & Train (s) & {Infer (ms)} & {Params.} & Train (s) & {Infer (ms)} & {Params.} \\
    \midrule
    {SharedBottom} & {9.83}& {3.18}& {3.43M} & {372.54}& {6.80}& {69.53M} & {440.18}& {6.38}& {12.35M} \\
    {MMoE} & {11.06}& {2.99}& {3.42M} & {398.51}& {8.63}& {69.51M} & {449.05}& {6.67}& {12.31M} \\
    {PLE} & {11.42}& {3.77}& {3.43M} & {370.02}& {9.46}& {69.81M} & 537.14 & {8.62}& {12.35M} \\
    {STAR} & {11.23}& {4.63}& {3.50M} & {355.32}& {9.21}& {69.90M} & {448.23}& {8.14}& {12.38M} \\
    {SAR-Net} & {10.08}& {4.08}& {3.44M} & {330.12}& {6.76}& {69.59M} & {410.71}& {6.52}& {12.31M} \\
    {M2M} & {18.02}& {9.01}& {3.54M} & {357.25}& {13.83}& {72.87M} & 553.64 & {11.71}& {12.38M} \\
    {AdaSparse} & {10.23}& {2.53}& {3.43M} & {331.01}& {5.79}& {69.79M} & 471.53 & {4.38}& {12.34M} \\
    {ADL} & {10.36}& {2.64}& {3.45M} & {358.30}& {4.83}& {69.56M} & 439.51 & {4.08}& {12.44M} \\
    {EPNet} & {10.03}& {3.02}& {3.43M} & {360.04}& {4.64}& {69.95M} & 450.68 & {4.33}& {12.30M} \\
    {PPNet} & {12.04}& {4.21}& {3.60M} & {380.04}& {5.31}& {70.54M} & 525.83 & {4.42}& {12.52M} \\
    {HAMUR} & 14.29& 7.68& 3.77M& 368.32& 7.65& 71.32M& 523.56& 7.81& 12.36M\\
    {M$^3$oE} & 13.56& 6.32& 3.42M& 364.25& 6.98& 69.36M& 478.63& 6.85& 12.21M\\

\bottomrule

    \end{tabular}}
    \vspace{-2mm}
    }
\end{table*}%
\section{Experiment}\label{sec:experiment}
This section presents the experimental results. We first describe the experimental setup, followed by the results analysis, including performance analysis, efficiency analysis, and scenario number analysis, as outlined below:

\subsection{Benchmarking Settings} \label{sec:experiment setting}
The experimental setup, including dataset processing, metrics used, and parameter configuration, is introduced below:
\begin{itemize}[leftmargin=*]
\item For each dataset, features are independently processed using discretization and bucketing techniques. These features are classified into three categories: sparse features (discretized attributes), dense features (continuous attributes), and scenario-specific features (operations specific to the scenario). The datasets are typically divided into training, evaluation, and testing sets in an 8:1:1 ratio, unless predefined splitting rules are specified.
\item For evaluation metrics, we follow methodologies from prior MSR works like~\cite{sheng2021one, yang2022adasparse,li2023hamur, wang2023plate, chang2023pepnet}, using Area Under the ROC Curve (AUC) and Logloss as metrics. Higher AUC or lower Logloss indicates better model performance.
\item For parameter settings, we ensure a fair comparison by configuring each model within a consistent search space and maintaining similar parameter magnitudes across datasets. All models we reproduced are carefully follow the original paper, besides, experiments are run 10 times with different random seeds to ensure the robustness of the results.
\end{itemize}
More details about scenario splitting information and experiment can be found in Appendix~\ref{sec_appendix:Scenario Information Analysis} and Appendix~\ref{sec_appendix:Experiment Settings}.

\subsection{Performance Analysis}\label{sec:comanalysis}

The overall results are presented in Table~\ref{tab:Performance}, and the analysis is shown as follows:

\begin{itemize}[leftmargin=*]

\item As shown in Table~\ref{tab:Performance}, models that incorporate an expert structure (e.g., MMoE, PLE, SAR-Net, M$^3$oE) generally outperform those that model different scenarios directly (e.g., SharedBottom, ADL). This suggests that expert-structured models are more effective at capturing complex inter-scenario dynamics at deeper network layers. Additionally, models capable of dynamically adjusting key structures or parameters based on varying scenarios (e.g., M2M, AdaSparse, HAMUR) outperform those with static expert structures, highlighting their ability to exert more precise control over the influence of hidden structures on scenario performance. This, in turn, enhances the understanding of scenario correlations and improves overall model performance. Furthermore, the size of the dataset does not appear to directly correlate with the performance disparity between models.

\item Additionally, we observe that variability in performance under sparse conditions—where user-item interactions are limited—has a significant impact on overall model effectiveness. Top-performing models consistently deliver strong results across all conditions, while less effective models tend to show improvements only in select cases. Notably, models leverage techniques such as collaborative-shared architectures (STAR) or meta-learning (M2M) to balance across domains, enhancing performance in sparse conditions without sacrificing effectiveness in more data-rich settings. This highlights the importance of capturing scenario correlations to mitigate the effects of sparsity and to promote unified performance gains across diverse environments.

\end{itemize}

\subsection{Efficiency Analysis}

We present the efficiency results, including training time, evaluation time, and parameter size for each model across different datasets in Table~\ref{tab:efficiency}. The analysis is as follows:
\begin{itemize}[leftmargin=*]
\item 
We observe that the models exhibited a range of parameter sizes, highlighting the trade-offs between model complexity and efficiency. For relatively small datasets, such as MovieLens and Douban, the training times were notably lower, reflecting the reduced computational load compared to the larger dataset, Ali-CCP. It is evident that model efficiency is influenced not only by algorithmic design but also significantly by the characteristics of the dataset, including the number and intrinsic nature of its features. This consideration is crucial for applications with limited computational resources. Across different models, the model sizes remained within the same order of magnitude, primarily because most parameters in recommender systems derive from embedding layers. Our findings underscore the importance of selecting the appropriate model based on both computational budget and the specific characteristics of the dataset. We believe these efficiency results provide a valuable reference for scholars aiming to select suitable models or datasets based on their resources in practical machine learning applications.
\end{itemize}

\subsection{Scenario Number Analysis} \label{sec:number analysis}
\begin{figure*}[t]
	\centering
		\includegraphics[width=0.9\linewidth]{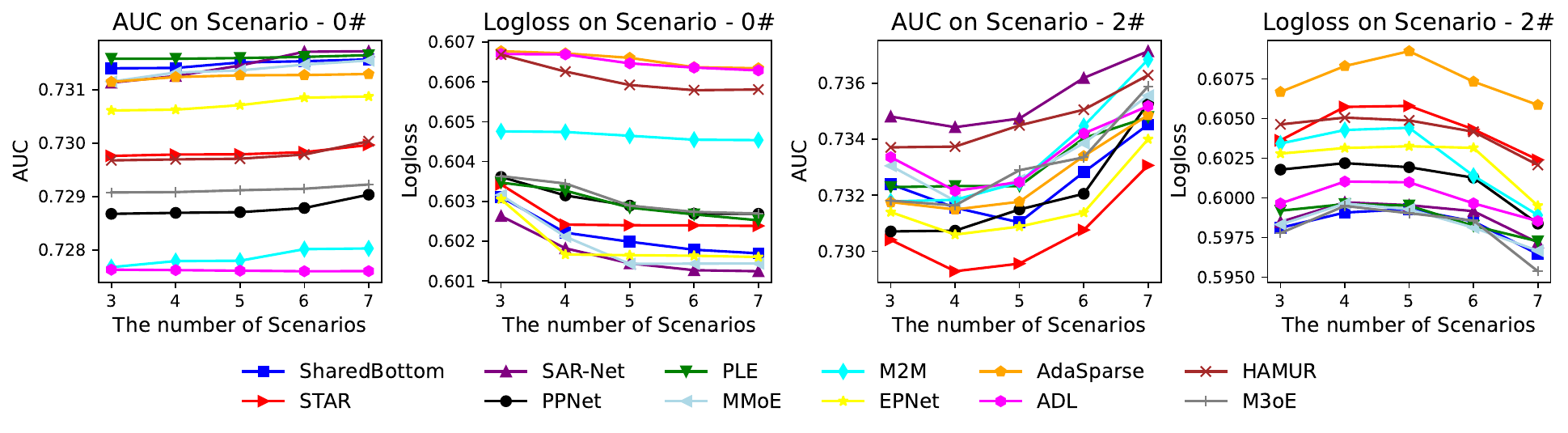}
\vspace{-1mm}
\caption{Performance versus number of scenarios on Scenario-0\# and Scenario-2\#.}
\vspace{-1mm}
\label{fig:scenarionum}
\end{figure*}

In MSR systems, there is a complex relationship between the number of scenarios and performance. To analyze this relationship, we use the KuaiRand dataset, varying the number of scenarios from 3 to 7, and observe performance in two scenarios: a dense scenario (Scenario-0\#), which contains more interactions, and a sparse scenario (Scenario-2\#), which contains fewer interactions. As shown in Figure~\ref{fig:scenarionum}, the scenario interaction number is shown in Table~\ref{Table:kuairand_domain_number_experiment_statistics}.

\begin{itemize}[leftmargin=*]

\item 
We observe that the performance in both scenarios improves as the number of scenarios increases from 3 to 7. This improvement can be attributed to the increased number of instances, which augment the dataset and enhance domain collaboration, thereby boosting overall performance. However, in the sparse Scenario-2\#, we observe a ``seesaw effect'', where an initial performance drop is followed by an improvement. This drop occurs because the addition of the sparse scenario negatively impacts overall performance, as seen in models such as SharedBottom, ADL, and STAR. Notably, SAR-Net demonstrates a strong ability to balance performance across both dense and sparse scenarios, maintaining consistent results. In practical deployments, it is crucial to balance the trade-off between performance fluctuations across multiple scenarios and adapt the model to specific conditions.

 \end{itemize}

\section{Industrial Experiment} \label{sec:Industrial Experiment}
The MSR task is highly relevant to real-world recommendation systems. Compared to public datasets, online multi-scenario settings are much more complex due to the larger number and greater diversity of scenarios, as well as the inclusion of a wider range of features. Furthermore, current public MSR datasets are not exclusively designed for MSR research. Therefore, to (1) validate the feasibility of our benchmark in practical industrial scenario settings and (2) provide a reliable benchmark for industrial applications, we tested our benchmark using an industrial dataset from an online advertising platform. This dataset includes 10 different scenarios and 108 features, spanning nine days. The first seven days are used for training, while the last two are reserved for validation and testing. It encompasses both common and scenario-specific user and item spaces. Supplemental information about the dataset can be found in Table~\ref{tab:Industrial dataset reference sheet}.

\begin{table}[ht]
\centering
\vspace{-2mm}
\caption{Industrial dataset reference sheet.}
\vspace{-2mm}
\label{tab:Industrial dataset reference sheet}
\setlength{\tabcolsep}{1mm}{
\scalebox{0.7}{
\begin{tabular}{l|p{8cm}}
\toprule
\textbf{Number of Features}   & \makebox[\linewidth]{108} \\ \midrule
\textbf{Number of Scenarios}  & \makebox[\linewidth]{10}  \\ \midrule
\textbf{Interaction}          & \makebox[\linewidth]{3M}  \\ \midrule
\textbf{Features Categories}              & 1. User features: attributes related to the user's profile and behavior, such as user city, click history, etc. \\
                                 & 2. App features: attributes related to the specific application or service being used, such as application category, application size, etc. \\
                                 & 3. Context features: context features that users interact with, such as device name, time, domain id, etc. \\ \midrule
\textbf{Train/Val/Test Splitting}         & \makebox[\linewidth]{7:1:1 (Split by days)}                                                                                      \\ \midrule
\textbf{Scenario Interaction}          & S-0: 301,654; S-1: 91,468; S-2: 22,986; S-3: 10,928; S-4: 316,734; S-5: 16,288; S-6: 383,791; S-7: 459,370; S-8: 87,353; S-9: 655,569 \\ 
\bottomrule
\end{tabular}
\vspace{-3mm}
}}
\end{table}

\subsection{Result Analysis}

Table~\ref{tab:indusperform} presents the results on the industrial dataset. Compared to other public datasets, this industrial dataset features a significantly larger number of scenarios and features. It is observed that M2M and M$^3$oE exhibit superior performance, demonstrating their ability to jointly handle a large number of scenarios. This finding is consistent with the observation in Table~\ref{tab:Performance}, where the public dataset Kuairand, which contains many more scenarios, also demonstrates great performance. This reveals that the innovative designs of meta cells and the multi-level fusion mechanism may lead to substantial improvements when dealing with real-world scenarios.

\begin{table}[ht]
\centering
\vspace{-2mm}
\caption{Performance comparison on the industrial dataset.}
\vspace{-2mm}
\label{tab:indusperform}%
\setlength{\tabcolsep}{1mm}{
\scalebox{0.8}{
    \begin{tabular}{c|cccccc}
    \toprule
     & \textbf{SharedBottom} & \textbf{MMoE}  & \textbf{PLE}   & \textbf{STAR}  & \textbf{SAR-Net} & \textbf{M2M} \\
    \midrule
    AUC   & 0.8276  & 0.8301  & 0.8330  & 0.8310  & 0.8355  & \textbf{0.8392} \\
    Logloss & 0.1521  & 0.1567  & 0.1496  & 0.1503  & 0.1528  & 0.1494\\
    \midrule
      & \textbf{AdaSparse} & \textbf{ADL} & \textbf{EPNet} & \textbf{PPNet} & \textbf{HAMUR} & \textbf{M$^3$oE} \\
     \midrule
     AUC & 0.8224  & 0.8358  & 0.8349  & 0.8318 & 0.8353 & 0.8384 \\
     Logloss & 0.1596  & \textbf{0.1489 } & 0.1517  & 0.1555 & 0.1501&  0.1492\\
    \bottomrule
    
    \end{tabular}%
    \vspace{-3mm}
  }}
\end{table}%

\subsection{Ethical Clarification}\label{sec: Ethical Concerns}
In utilizing the industrial dataset, we prioritize ethical considerations, particularly user privacy protection and responsible data usage. To ensure data privacy, we implement comprehensive safeguarding measures: all user-specific identifiers are removed to prevent sensitive data leakage; demographic attributes such as gender and location are transformed into numerical features through irreversible hashing; and behavioral data is similarly anonymized with explicit user consent obtained prior to collection. The dataset contains only explicit user interactions like clicks, excluding more personal engagement metrics such as favorites, likes, and comments. Our data collection process strictly adheres to all relevant legal and regulatory requirements, with all data gathered from a single online platform under user authorization and signed consent. No data is collected from users who have not provided explicit consent.
\section{Conclusion}
This paper introduces \name, the first benchmark tailored for multi-scenario recommendation (MSR) systems. \name provides a standardized, reproducible framework for evaluating diverse MSR models and promotes knowledge sharing within the research community. We contribute in three key ways: (1) \name enables systematic model comparisons and drives progress in MSR research; (2) it offers a complete pipeline, covering data processing, training, evaluation, logging, and open sourcing, to enhance transparency and reproducibility; and (3) it reproduces twelve representative MSR models across seven datasets, supporting robust evaluation and experimentation. Looking ahead, we plan to explore the integration of LLMs into MSR applications~\cite{fu2023unified, wang2024llm4msr, luo2025tapo, li2024syneg, gao2025samplellm, zhang2025process, liu2024moe, fu2025training}.

\begin{acks}
This research was partially supported by National Natural Science Foundation of China (No.62502404), Hong Kong Research Grants Council's Research Impact Fund (No.R1015-23), Collaborative Research Fund (No.C1043-24GF), General Research Fund (No.11218325), Institute of Digital Medicine of City University of Hong Kong (No.9229503), and Huawei (Huawei Innovation Research Program).
\end{acks}
\appendix

\section{Scenario Information Analysis} \label{sec_appendix:Scenario Information Analysis}
\subsection{Scenario Splitting Strategy} \label{sec_appendix:Scenario Splitting Strategy}

Unlike traditional CTR prediction tasks, MSR models require scenario-unified prediction, necessitating a scenario indicator in the dataset to enable scenario splitting. The domain indicator thus becomes essential for distinguishing scenarios. Scenario-splitting strategies generally fall into three categories:

\begin{itemize}[leftmargin=*]
\item \textbf{Context Feature Splitting}: Uses predefined context features to distinguish scenarios, such as ad area, page number, or position. For example, Ali-CCP and KuaiRand use ``Tab'' (page number) and ``301'' (position) for segmentation.

\item \textbf{Item Feature Splitting}: Differentiates scenarios based on item types. In Amazon, scenarios are split by product category; in Douban, by platform name.

\item \textbf{User Feature Splitting}: Segments scenarios by user attributes. In MovieLens, for instance, interactions are grouped by user age.
\end{itemize}

Recent work~\cite{guo2023dffm, jia2024d3} explores automated scenario splitting based on data-driven characteristics, though this area remains relatively underexplored.

\subsection{Scenario Analysis}
In this section, we analyze scenario distributions across various datasets.
\begin{itemize}[leftmargin=*]

\item We assess distribution uniformity using the Coefficient of Variation (COV), where higher values indicate greater imbalance (Table~\ref{tab:interdomain}). KuaiRand exhibits the most uneven distribution due to user concentration on the homepage, while MovieLens shows the most uniform distribution with scenarios evenly split by age. Douban skews toward movies due to frequent browsing behavior, and Ali-CCP's COV of approximately 0.9 indicates similarly unbalanced scenario distribution. Mind and Amazon demonstrate more balanced distributions, reflected in their lower COV values.

\item We examine scenario intersections to understand user-item overlap (Table~\ref{tab:interdomain}). Industrial dataset intersections remain unavailable due to privacy constraints. MovieLens user groups share most movies while maintaining distinct preferences. KuaiRand reveals bimodal user distribution with long-tail item patterns—Scenarios 3 and 4 share 704 of 832 users but differ in item interactions. Ali-CCP's Scenario 1 represents only 1\% of interactions, creating skewed distribution and minimal overlap. For Amazon, Douban, and Mind—which lack explicit scenario features—we apply dataset-specific segmentation strategies. Amazon scenarios by item type show large user overlap with evenly distributed interactions. Douban's platform-based split (Book, Music, Movie) demonstrates movie dominance, though over 1,000 users span all three platforms. Mind's news category segmentation reveals over 600,000 users overlapping across feeds.

\end{itemize}

\begin{table}[t]
  \centering
  \vspace{-2mm}
  \caption{Dataset statistics for scenario intersection.}
  \vspace{-2mm}
  \scalebox{0.7}{
    \begin{tabular}{c|c|c|c|c}
    \toprule
    Dataset& COV & Scenario Indicator  & \# User Intersection & \# Item Intersection  \\
    \midrule
    \multirow{3}{*}{MovieLens}&\multirow{3}{*}{0.3186} & S-0~$\cap$~S-1 & -    &  3,320   \\
    &&S-1~$\cap$~S-2 & -     &  3,448   \\
    &&S-0~$\cap$~S-2 & -     &  3,354   \\
    \midrule
    \multirow{6}{*}{KuaiRand} &\multirow{6}{*}{1.3552} & S-0~$\cap$~S-1 & 961	    &  380,375   \\
    &&S-0~$\cap$~S-2 & 160	     &  64,292   \\
    &&S-1~$\cap$~S-2 & 162	     &  213,106   \\ 
    &&S-1~$\cap$~S-3 & 832	     &  264,931   \\
    &&S-2~$\cap$~S-3 & 141	     &  66,063   \\ 
    &&S-3~$\cap$~S-4 & 704	     &  2,721   \\ 
    \midrule
    \multirow{3}{*}{Ali-CCP} &\multirow{3}{*}{0.9180} & S-0~$\cap$~S-1 & 814	    &  188,510   \\
    &&S-1~$\cap$~S-2 & 515	     &  188,590   \\
    &&S-0~$\cap$~S-2 & 2,385	     &  465,694   \\
    \midrule
    \multirow{3}{*}{Amazon} &\multirow{3}{*}{0.2696} & S-0~$\cap$~S-1 & 4,220	    &  -   \\
    &&S-1~$\cap$~S-2 & 6,557	     &  -  \\
    &&S-0~$\cap$~S-2 & 7,026	     &  -  \\
    \midrule
    \multirow{3}{*}{Douban} &\multirow{3}{*}{1.1053}
    &S-0~$\cap$~S-1 &  1,736   &  -   \\
    &&S-1~$\cap$~S-2 & 1,815	 &  -   \\
    &&S-0~$\cap$~S-2 & 2,209	 &  -   \\
    \midrule
    \multirow{6}{*}{Mind} &\multirow{6}{*}{0.5611}
    &S-0~$\cap$~S-1  & 675,343  & -    \\
    &&S-1~$\cap$~S-2 & 646,049	 & -    \\
    &&S-2~$\cap$~S-3 & 633,042	 & -    \\
    &&S-0~$\cap$~S-2 & 689,568	 & -    \\
    &&S-1~$\cap$~S-3 & 626,604	 & -    \\
    &&S-0~$\cap$~S-3 & 653,595	 & -    \\
    % \midrule
    % \multirow{3}{*}{Tenrec} &\multirow{3}{*}{0.8413}
    % &S-0~$\cap$~S-1 & 987,743    &  -   \\
    % &&S-1~$\cap$~S-2 & 454,158	     &  -   \\
    % &&S-0~$\cap$~S-2 & 455,221	     &  -   \\
    \bottomrule
    \end{tabular}%
    }
    \vspace{-2mm}
  \label{tab:interdomain}%
\end{table}%

\section{Experiment Settings} \label{sec_appendix:Experiment Settings}
% \subsection{More Implementation Details} \label{sec_appendix:More Implementation Details}
In this part, we present the experiment setting during our experiment. Our framework is implemented using PyTorch. Empirically, we set the feature embedding dimension $d$ to 16. We customized batch sizes for each dataset: 4096 for MovieLens, Amazon, Douban and Mind, 9,048 for both Kuairand and the industrial dataset, and 102,400 for Aliccp. Experiments were conducted on a single GPU of Tesla V100 PCIe 32GB, utilizing the Adam optimizer. The initial learning rate was set to 1e-3. To enhance training performance, we incorporated an early stopping strategy and a learning rate scheduler for optimal adjustment.

\begin{table}[t]
  \centering
  
  \caption{Scenario distribution for scenario-number experiments.}
  \vspace{-2mm}
  \label{Table:kuairand_domain_number_experiment_statistics}
  \scalebox{0.9}{
    \begin{tabular}{cccc}
    \toprule
    Scenario & \# Interaction & Scenario & \# Interaction \\
    \midrule
    Scenario 0 & 7,760,237 & Scenario 4 & 183,403\\
    Scenario 1 & 2,407,352 & Scenario 5 & 37,418 \\
    Scenario 2 & 895,385 & Scenario 6 & 17,430 \\
    Scenario 3 & 402,366 & - & - \\
    \bottomrule
    \end{tabular}%
    }
    \vspace{-2mm}
\end{table}

\section{GenAI Usage Disclosure}

In this study, we used generative large language models solely for writing assistance (e.g., typographical correction). No LLM-based techniques were employed in any other part of the work.

%%
%% The next two lines define the bibliography style to be used, and
%% the bibliography file.
\bibliographystyle{ACM-Reference-Format}
\balance
\bibliography{bibfile}

\end{document}